\documentclass[showpacs,prl,aps,10pt,superscriptaddress,altaffilletter,floatfix,twocolumn]{revtex4-1}
\usepackage{graphicx}
\usepackage{hyperref}
\usepackage{amsmath}
\usepackage{amssymb}
\usepackage{color}
\usepackage{xspace}
\usepackage{dcolumn}
\usepackage{soul}
\newcolumntype{C}[1]{>{\centering\let\newline\\\arraybackslash\hspace{0pt}}m{#1}}

\begin{document}

\newcommand{\isot}[2]{$^{#2}$#1}
\newcommand{\isotbold}[2]{$^{\boldsymbol{#2}}$#1}
\newcommand{\xeiso}{\isot{Xe}{136}\xspace}
\newcommand{\thsrc}{\isot{Th}{228}\xspace}
\newcommand{\cosrc}{\isot{Co}{60}\xspace}
\newcommand{\rasrc}{\isot{Ra}{226}\xspace}
\newcommand{\cssrc}{\isot{Cs}{137}\xspace}
\newcommand{\betascale}  {$\beta$-scale}
\newcommand{\kevkgyr}  {keV$^{-1}$ kg$^{-1}$ yr$^{-1}$}
\newcommand{\nonubb}  {$0\nu \beta\!\beta$\xspace}
\newcommand{\nonubbbf}  {$\boldsymbol{0\nu \beta\!\beta}$\xspace}
\newcommand{\twonubb} {$2\nu \beta\!\beta$\xspace}
\newcommand{\bb} {$\beta\!\beta$\xspace}
\newcommand{\vadc} {ADC$_\text{V}$}
\newcommand{\uadc} {ADC$_\text{U}$}
\newcommand{\mus} {\textmu{}s}
\newcommand{\chisq} {$\chi^2$}
\newcommand{\mum} {\textmu{}m}
\newcommand{\red}[1]{{\xspace\color{red}#1}}
\newcommand{\blue}[1]{{\xspace\color{blue}#1}}
\newcommand{\RunTwoA}{Run 2a}
\newcommand{\RunTwo}{Run 2}
\newcommand{\RunTwoBC}{Runs 2b and 2c}
\newcommand{\SP}[1]{\textsuperscript{#1}}
\newcommand{\SB}[1]{\textsubscript{#1}}
\newcommand{\SPSB}[2]{\rlap{\textsuperscript{#1}}\SB{#2}}
\newcommand{\pmasy}[3]{#1\SPSB{$+$#2}{$-$#3}}
\newcommand{\matel}{$M^{2\nu}$}
\newcommand{\psfac}{$G^{2\nu}$}
\newcommand{\tbeta}{T$_{1/2}^{0\nu\beta\beta}$}
\newcommand{\exolimit}[1][true]{\pmasy{2.6}{1.8}{2.1}$ \cdot 10^{25}$}
\newcommand{\exomeasurement}{\tbeta{}= \exolimit{}~yr}
\newcommand{\U}{\text{U}}
\newcommand{\V}{\text{V}}
\newcommand{\X}{\text{X}}
\newcommand{\Y}{\text{Y}}
\newcommand{\Z}{\text{Z}}
\newcommand{\bqcm}{${\rm Bq~m}^{-3}$}
\newcommand{\nonunorm}{N_{{\rm Err, } 0\nu\beta\beta}}
\newcommand{\nonunum}{n_{0\nu\beta\beta}}
\newcommand{\cussim}[1]{$\sim$#1}
\newcommand{\halflife}[1]{$#1\cdot10^{25}$~yr}
\newcommand{\numspec}[3]{$N_{^{#2}\mathrm{#1}}=#3$}
\newcommand{\TD}[1]{\textcolor{red}{#1}}
\newcommand{\PI}{Phase~I\xspace}
\newcommand{\PII}{Phase~II\xspace}
\newcommand{\Rn}{radon\xspace}
\newcommand\Tstrut{\rule{0pt}{2.6ex}} 

\setstcolor{blue}

\title{Search for Neutrinoless Double-Beta Decay with the Complete EXO-200 Dataset}

\author{G.~Anton}\affiliation{Erlangen Centre for Astroparticle Physics (ECAP), Friedrich-Alexander-University Erlangen-N\"urnberg, Erlangen 91058, Germany}
\author{I.~Badhrees}\altaffiliation{Permanent position with King Abdulaziz City for Science and Technology, Riyadh, Saudi Arabia}\affiliation{Physics Department, Carleton University, Ottawa, Ontario K1S 5B6, Canada}
\author{P.S.~Barbeau}\affiliation{Department of Physics, Duke University, and Triangle Universities Nuclear Laboratory (TUNL), Durham, North Carolina 27708, USA}
\author{D.~Beck}\affiliation{Physics Department, University of Illinois, Urbana-Champaign, Illinois 61801, USA}
\author{V.~Belov}\affiliation{Institute for Theoretical and Experimental Physics named by A.I. Alikhanov of National Research Centre ``Kurchatov Institute'', 117218, Moscow, Russia}
\author{T.~Bhatta}\affiliation{Department of Physics, University of South Dakota, Vermillion, South Dakota 57069, USA}
\author{M.~Breidenbach}\affiliation{SLAC National Accelerator Laboratory, Menlo Park, California 94025, USA}
\author{T.~Brunner}\affiliation{Physics Department, McGill University, Montreal H3A 2T8, Quebec, Canada}\affiliation{TRIUMF, Vancouver, British Columbia V6T 2A3, Canada}
\author{G.F.~Cao}\affiliation{Institute of High Energy Physics, Beijing 100049, China}
\author{W.R.~Cen}\affiliation{Institute of High Energy Physics, Beijing 100049, China}
% ...

\collaboration{EXO-200 Collaboration}

\date{\today}

\begin{abstract}

  A search for neutrinoless double-beta decay (\nonubb) in
  $^{136}$Xe is performed with the full EXO-200 dataset using
  a deep neural network to discriminate between \nonubb and
  background events. Relative to previous analyses, 
  the signal detection efficiency has been raised from $80.8\%$
  to $96.4\pm3.0\%$ and the energy resolution of the detector 
  at the Q-value of $^{136}$Xe \nonubb
  has been improved from $\sigma/E=1.23\%$ to $1.15\pm0.02\%$ with the upgraded detector. 
  Accounting for the new data, the median 90\% confidence level \nonubb
  half-life sensitivity for this analysis is
  $5.0 \cdot 10^{25}$~yr with a total $^{136}$Xe exposure of 234.1~kg$\cdot$yr. 
  No statistically significant evidence for \nonubb is observed, leading to a lower
  limit on the \nonubb half-life of $3.5\cdot10^{25}$~yr at the 90\% confidence
  level.

\end{abstract}


\maketitle

Double-beta decay is a second-order weak transition
in which two neutrons simultaneously decay into two protons.
While the mode with emission of two electrons and two antineutrinos (\twonubb)
has been observed in several nuclides in which single-beta
decay is suppressed~\cite{PDG18}, the 
hypothetical neutrinoless mode (\nonubb)~\cite{PhysRev.56.1184}
is yet to be discovered. 
The search for \nonubb is recognized as the most sensitive probe for
the Majorana nature of neutrinos~\cite{Rodejohann:2011mu,Vergados:2012xy,DellOro:2016tmg,Dolinski:2019nrj}. 
Its observation would provide direct evidence for a 
beyond-the-Standard-Model process that violates lepton number conservation,
as well as help constrain the absolute mass scale of neutrinos~\cite{Engel:2016xgb}.

Recent experiments probing
a number of nuclides~\cite{EXO_PRL17,KamLAND-Zen:2016pfg,MD_17,Gerda2018,CUORE_First} 
have set lower limits on the \nonubb half-life
with sensitivities in the range 
$10^{25} - 10^{26}$~yr at 90\% confidence level (CL). 
Exploiting the advantages of a liquid xenon (LXe) cylindrical 
time projection chamber (TPC) filled with LXe enriched to 80.6\% in \xeiso~\cite{Albert2013}, 
EXO-200~\cite{Auger:2012gs} achieved a sensitivity of 
$3.7 \cdot 10^{25}$~yr~ with the most recent \nonubb search~\cite{EXO_PRL17}, 
while the most sensitive search to date for the same isotope reached
$5.6 \cdot 10^{25}$~yr~\cite{KamLAND-Zen:2016pfg}.
Here we report on a search with similar sensitivity to the 
previous best search.

In Dec.~2018, EXO-200 completed data taking with the upgraded detector 
(``\PII'', May~2016 to Dec.~2018), 
after collecting an exposure similar to that of its first run 
(``\PI'', Sept.~2011 to Feb.~2014).
This letter reports a search for \nonubb using the full EXO-200 dataset, 
which after data quality cuts~\cite{Albert2013} totals 1181.3~d of livetime.
This represents 
approximately a 25\% increase in exposure relative to 
the previous search~\cite{EXO_PRL17} that already included nearly half of the \PII dataset.
In addition to the new data
acquired between Jun.~2017 and Oct.~2018, 
this search introduces several analysis developments 
to optimize the detector sensitivity to \nonubb,
including the incorporation of a deep neural network (DNN) to discriminate
between background and signal events.

In the EXO-200 detector, 
a common cathode splits the LXe TPC into
two drift regions, 
each with radius $\sim$18~cm and drift length $\sim$20~cm. 
The TPC is enclosed by a radiopure thin-walled copper vessel.
The electric field in the drift regions was raised from 380~V/cm in \PI
to 567~V/cm in \PII to improve the energy resolution, 
since it was found that, after the detector was re-started, 
higher voltage values on the cathode were stable. The
ionization produced from interactions in the LXe is read out
after being drifted to crossed-wire planes at each anode, 
while the scintillation light produced at the interaction time
is collected by arrays of large area avalanche photodiodes (LAAPDs)~\cite{Neilson:2009kf}
located behind the wire planes. 

The underground location of the experiment,
the Waste Isolation Pilot Plant (WIPP) near Carlsbad New Mexico, 
provides an overburden
of 1624$^{+22}_{-21}$ meters of water equivalent~\cite{EXO_cosmogenics}.
In addition to several layers of passive shielding,
including $\sim$50 cm of HFE-7000 cryofluid~\cite{m3m},
5.4~cm of copper and
$\sim$25~cm of lead in all directions~\cite{Auger:2012gs},
an active muon veto system with scintillator panels on four sides 
allows prompt identification of $>94\%$ of the cosmic ray muons
passing through the TPC. 
This system is also used 
in this analysis to reject background events 
arising from cosmogenically produced \isot{Xe}{137}, which primarily 
decays via $\beta$ emission with a half-life of 3.82~min~\cite{EXO_PRL17,Xe137_HL}.

Each TPC event is reconstructed by grouping charge and light signals 
into individual energy deposits.
Ionization signals measured by two wire planes, an induction plane
(V-wires) and a collection plane (U-wires),
provide information about the coordinates $x$ and $y$ perpendicular to the drift field.
The $z$ position, along the drift direction, is obtained from the time delay
between the prompt light and the delayed charge signals together with the
measured electron drift velocity~\cite{EXO_diffusion}.
Events reconstructed with single and multiple energy deposit(s) are referred 
to as ``single-site'' (SS) and ``multi-site'' (MS).
\nonubb events are predominantly SS whereas backgrounds are mostly MS.
While the main procedures for spatial reconstruction
are the same as in previous searches~\cite{Albert2013,EXO_Nature,EXO_PRL17},
the \nonubb detection efficiency has been raised to
$97.8\pm3.0$\% ($96.4\pm3.0$\%) in \PI (\PII) 
from $82.4\pm3.0$\% ($80.8\pm2.9$\%) \cite{EXO_PRL17}
by relaxing two selection criteria. 
First, 
the time required for events to be separated from all other reconstructed
events has been reduced from $>1$~s to $>0.1$~s.
This time cut is still at least two orders of magnitude longer 
than expected from typical time-correlated backgrounds 
seen in the detector~\cite{EXOBkgd2013,EXO_cosmogenics}, while the 
\nonubb efficiency loss 
due to accidental coincidence is
reduced from 7\% to 0.5\%.
Second, the search presented here includes events containing deposits
without a detected V-wire signal 
if these deposits contribute $<40\%$ of the total event energy, 
which were removed in the previous analyses. 
Because of the higher energy threshold for signal detection on the V-wires
($\sim$200~keV) versus the U-wires ($\sim$90~keV),
a significant number of events with small energy deposits
are well-reconstructed by
the U-wires but incompletely on the V-wires,
resulting in events with full $z$ reconstruction 
but incomplete $xy$ reconstruction for smaller energy deposits.
Relaxing this 3D-cut criterion only recovers MS events and retrieves
almost all potential \nonubb events with incomplete $xy$ reconstruction due to small, 
separated energy deposits from bremsstrahlung.
While \nonubb primarily induces SS events,
the smaller fraction of MS \nonubb events 
can be distinguished from the dominant $\gamma$ backgrounds 
using a discriminator for MS events (described below),
resulting in an enhancement in the \nonubb half-life sensitivity.

Events within the fiducial volume (FV) are required to lie within
a hexagon in the $xy$ plane with apothem of 162~mm.
They are further required to be more than 10~mm away
from the cylindrical PTFE reflector, 
as well as the cathode and the V-wire planes.
This FV contains $3.31\cdot10^{26}$ atoms of $^{136}$Xe,
with an equivalent mass of 74.7~kg.
While the incomplete $xy$-matched energy deposits may fall outside the FV,
this effect is determined by detector simulations 
to have a negligible effect on 
the estimated detection efficiency due to
the energy requirements imposed on these events.
The $^{136}$Xe exposure of the entire dataset after data quality 
cuts and accounting for livetime loss due to vetoing events coincident
with the muon veto
is 234.1~kg$\cdot$yr, or 1727.5~mol$\cdot$yr,
with 117.4 (116.7)~kg$\cdot$yr in \PI (\PII).

The detector response to \nonubb decays and background interactions is modeled by 
a detailed Monte Carlo (MC) simulation based on GEANT4~\cite{GEANT42006}.
This MC simulation models the energy deposits 
produced by interactions in the LXe, then propagates the ionization 
through the detector to produce waveforms associated to readout channels. 
These simulated waveforms are input
to the same reconstruction and analysis framework used for data waveforms.
Calibration data with external $\gamma$ sources located 9 (11) cm away from the FV
at set positions 
around the cathode (behind the anodes)~\cite{Albert2013} were regularly taken 
to validate the analysis. 

\begin{figure}[t]
\centering
\includegraphics[width=\columnwidth]{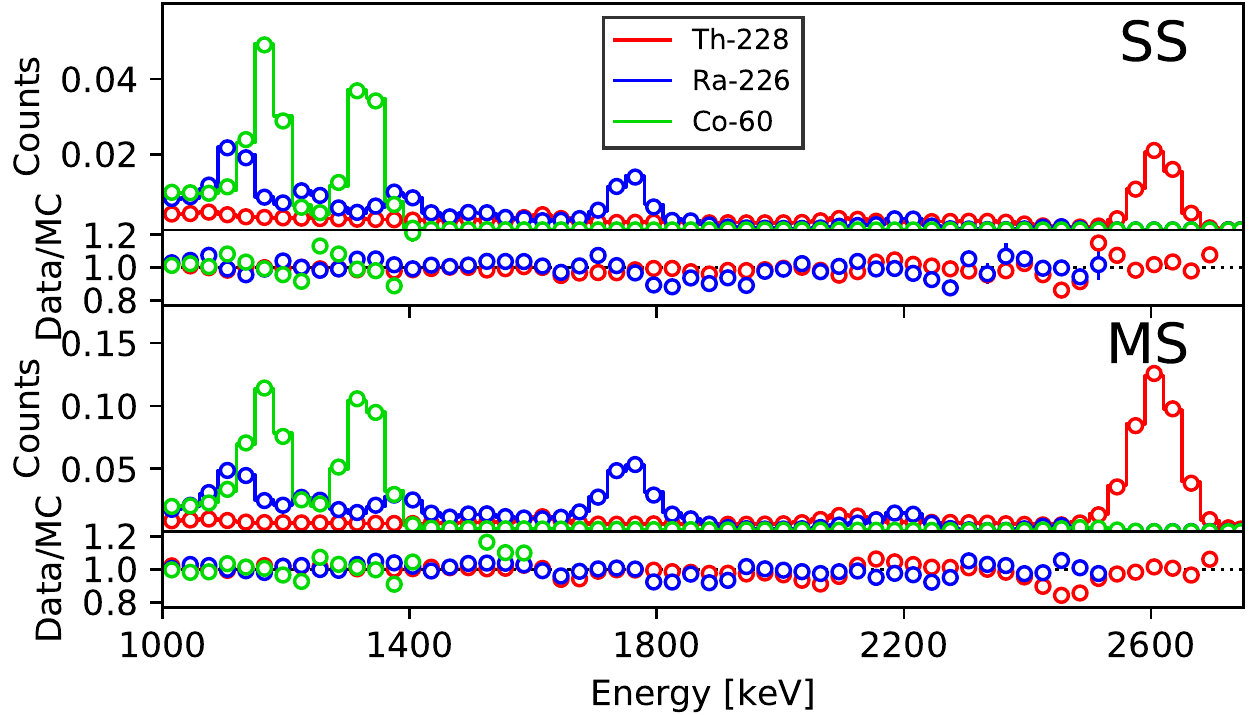}
\caption{
  Comparison of energy distributions in data (circles) and MC (lines)
  for SS (top half) and MS (bottom half) events from calibration sources positioned near the cathode. 
}
\label{fig:energy_source_agree}
\end{figure}

After the previous EXO-200 \nonubb search~\cite{EXO_PRL17},
a small fraction of the observed candidate events presented light-to-charge ratios
that were not fully consistent with their expected distributions.
Using calibration and \twonubb data, the distribution of the light-to-charge
ratio is measured and found to be approximately gaussian around the mean ratio.
While keeping the maximal search sensitivity, a cut is imposed requiring that events 
are within 2.5$\sigma$ of the mean of the distribution.
This improves the previous cut~\cite{Albert2013},
primarily aimed at removing $\alpha$ decays,
by also removing poorly reconstructed $\beta$ and $\gamma$ events
with an anomalous light-to-charge ratio.
All systematic errors associated with the signal detection efficiency 
are summarized in Tab.~\ref{tab:shape-syst}.

\begin{figure}[t]
\centering
\includegraphics[width=\columnwidth]{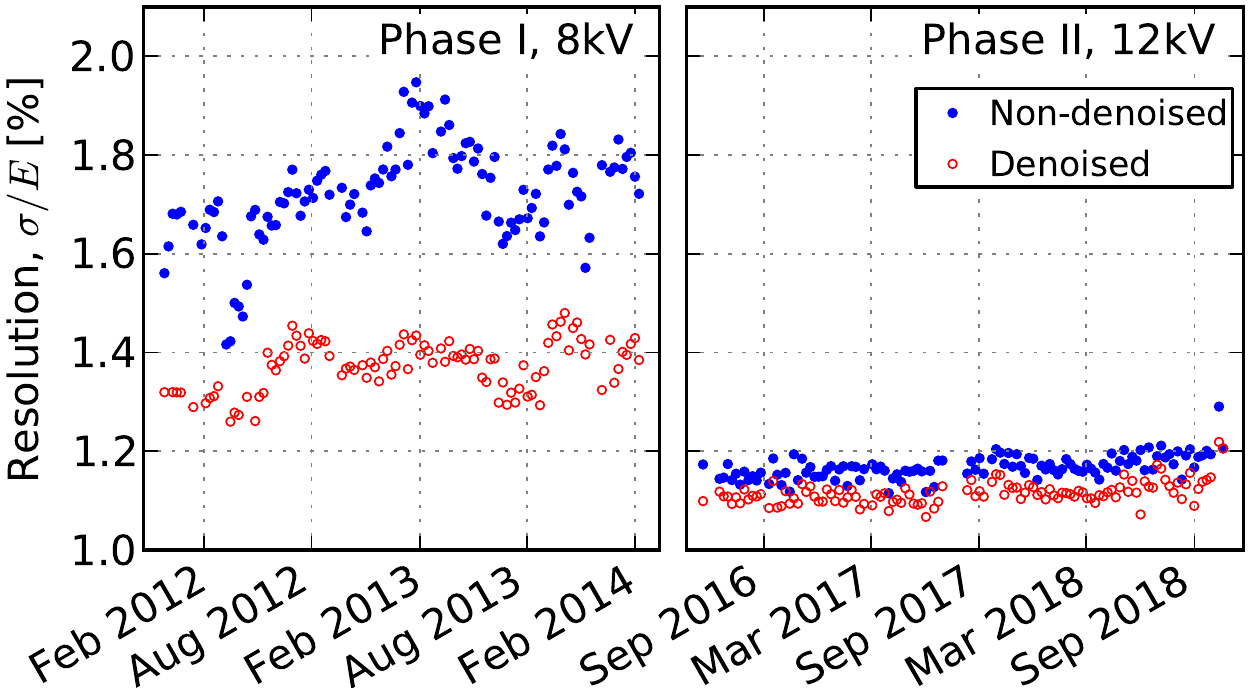}
\caption{
  Energy resolution
  of SS events
  measured using a \isot{Th}{228} calibration source
  deployed to a position near the cathode. 
  The effect of the de-noising algorithm
  and weekly variation of the resolution 
  at the 2615~keV $^{208}$Tl $\gamma$ line are 
  shown for both \PI and \PII.
  The resolution worsened slightly after a xenon recovery in July 2017
  due to a power outage. A degraded resolution due 
  to an increase of excess noise is visible in last weeks of \PII.
}
\label{fig:resolution}
\end{figure}

The reconstructed energy is determined by combining the anti-correlated 
charge and light signals~\cite{Conti2003}
to optimize the resolution at the \nonubb decay energy 
of $Q_{\beta\!\beta} = 2457.83\pm0.37$~keV~\cite{Redshaw:2007}.
An offline de-noising algorithm~\cite{EXO_denoising}, previously
introduced to account for excess APD read-out noise observed 
in \PI, has been further optimized with measurements of
the light response of the detector and adapted for \PII data. 
In addition, a proper modeling of mixed signals from the induced and 
collected charge in wires is introduced to the signal finder 
in the event reconstruction process.
The resulting energy measurement shows good spectral agreement between
data and simulation for SS and MS events using \thsrc, \rasrc and $^{60}$Co
calibration sources as shown in Figure~\ref{fig:energy_source_agree}.
The electronics upgrade carried out before \PII data taking resulted in 
substantially improved resolution and stability, as illustrated
in Figure~\ref{fig:resolution}. 
The average detector resolution is determined by uniformly
weighting all calibration data from several positions
and accounting for the detector livetime.
The averages for \PI and \PII are 
$\sigma/E\left(Q_{\beta\!\beta}\right)=1.35\pm0.09$\% and $1.15\pm0.02$\%, 
respectively. 

All data, including those previously reported, were blinded
to hide all candidate \nonubb SS events
having energy within $Q_{\beta\!\beta}\pm2\sigma$.
No information about such events is used in the
development of the techniques for this analysis.
New background discriminators are studied to
optimize the sensitivity of this search, while minimizing
the systematic errors.
The search for \nonubb is performed with a simultaneous 
maximum-likelihood (ML) fit to
the SS and MS energy spectra, with the discriminators
added as additional fit dimensions.
While \PI and \PII are fit independently
and then combined by summing their individual profile likelihoods for various signal hypotheses,
both use the same background model developed 
in~\cite{EXO_PRL17} composed of 
decays originating in the detector and surrounding materials.
Systematic errors are 
included in the ML fit as nuisance parameters
constrained by normal distributions.  
The median 90\% CL sensitivity is estimated using 
toy datasets (simulated trial experiments) generated from the 
MC probability density functions (PDFs)
of the background model.

\begin{figure}[t]
\centering
\includegraphics[width=\linewidth]{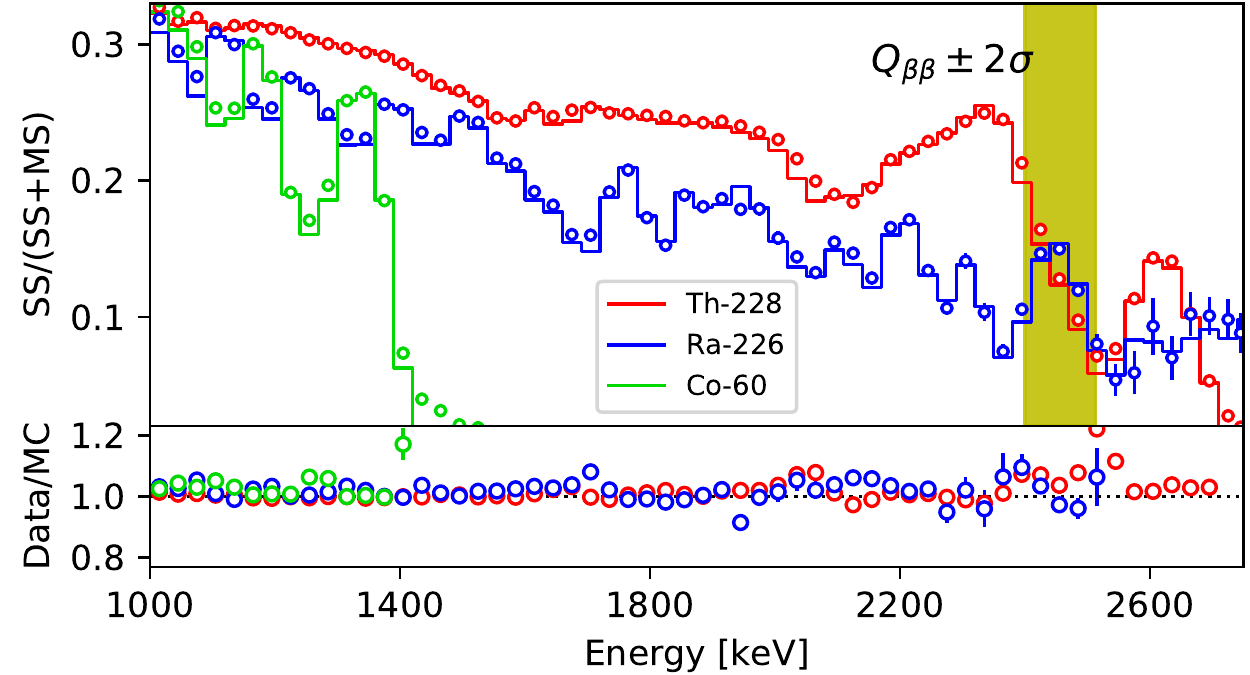}
\caption{
   SS fractions
   for MC (lines) and 
   data (circles) in \PII using 
   calibration sources positioned near the cathode. 
}
\label{fig:ss_fraction}
\end{figure}

The primary topological discrimination of backgrounds 
is the SS/MS event classification. Figure~\ref{fig:ss_fraction} shows the agreement
between source calibration data and MC  
for the ``SS fraction,'' SS/(SS+MS).
Because the relaxed 3D-cut recovers MS events,
the SS fraction near  $Q_{\beta\!\beta}$ is lowered from 
24\% (23\%) to 12\% (14\%) for the \thsrc (\rasrc) source
compared to previous searches.
Systematic errors related to the SS fractions are determined by 
comparisons between data and MC. Taking into account 
different calibration sources at various positions, these systematics 
are evaluated to be 5.8\% (4.6\%) for \PI (\PII).

Motivated by the results in~\cite{EXO_DL}, this analysis introduces
a new discriminator for SS and MS events using 
a DNN that relies 
on the waveforms of U-wire signals and is found to outperform the 
searches in~\cite{EXO_Nature,EXO_PRL17}.
The training inputs for the DNN are greyscale images built by 
arranging neighboring channels next to each other 
and encoding the amplitudes of U-wire waveforms as pixel values.
The training data is produced
in MC simulation for two classes of events in equal weights: 
background-like, composed of $\gamma$ events with uniform energy
distribution between 1000~keV and 3000~keV; and signal-like
\nonubb events with a random decay energy restricted to the same energy limits.
The location of the simulated interactions of both types are 
drawn uniformly from
the detector volume to focus discrimination only on the 
topological event characteristics. 
This dataset is divided into 90\% for training and 10\% for validation.
The DNN architecture is inspired by the Inception architecture proposed 
by Google~\cite{GoogleRef} and implemented with the Keras library~\cite{keras}
using the Tensorflow backend~\cite{tensorflow}.

\begin{figure}[t]
\centering
\includegraphics[width=\linewidth]{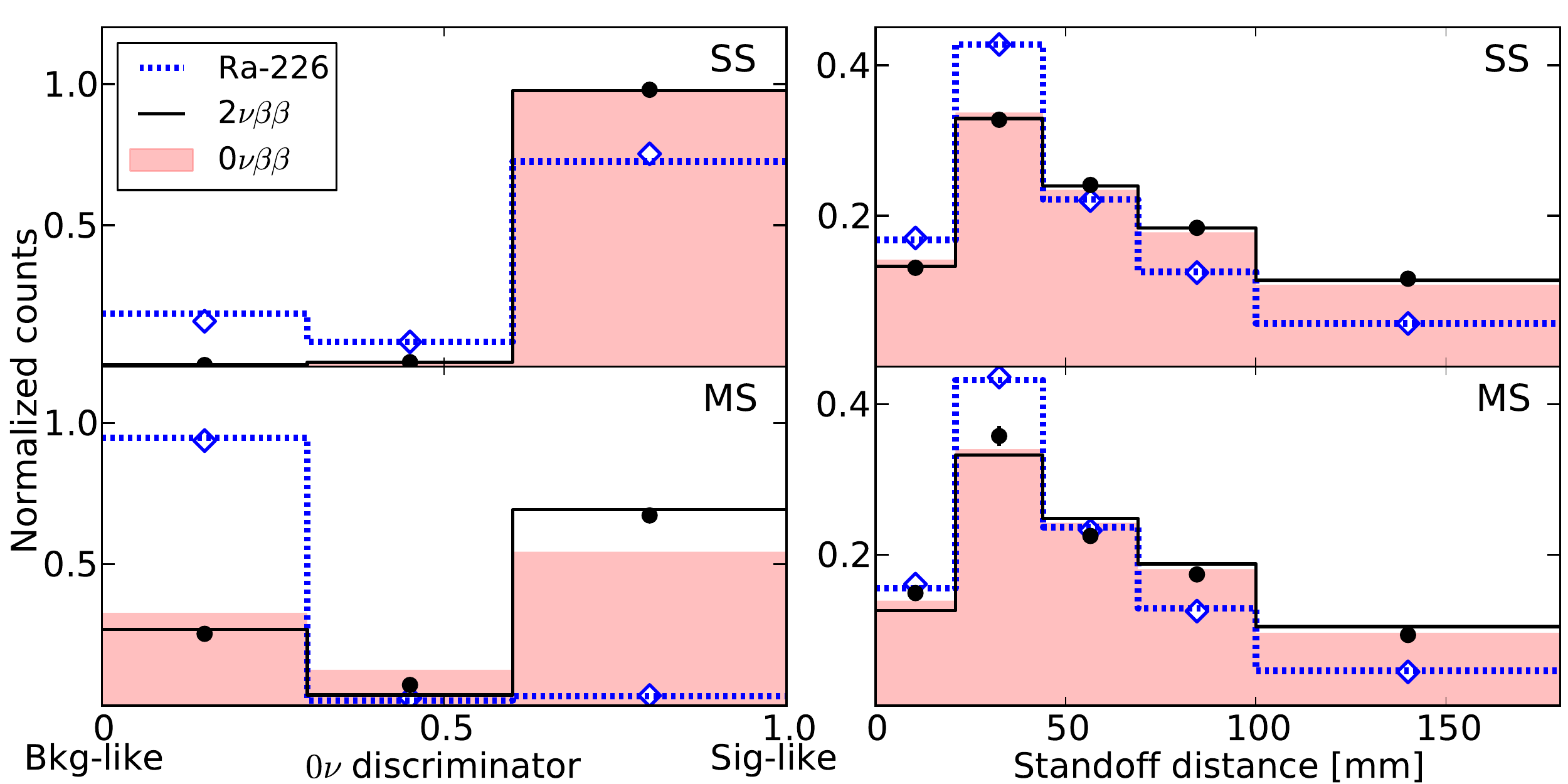}
\caption{
  Comparison between data (dots) and MC (solid/dashed
  lines) for the DNN \nonubb discriminator (left) and standoff distance (right).
  Shown are the distributions from the \rasrc calibration source (blue)
  and the background-subtracted \twonubb spectrum
  from low background data (black).  
  The simulated distributions for \nonubb events are indicated by the red filled region.
  The difference in DNN distribution between \nonubb and \twonubb events in MS is due to
  the higher rate of bremsstrahlung at higher electron energy. 
}
\label{fig:source_agree_dnn}
\end{figure}

The agreement for the DNN discriminator between data and MC is improved
when signals from U-wire waveforms are first identified by the signal finder in the 
EXO-200 reconstruction framework, and then used to re-generate the 
images. Since there is no spatial dependence in training the DNN 
for signal- and background-like events, 
the standoff distance 
(minimum distance between the event position and the 
closest material surface excluding the cathode),
is incorporated in the search as a 
third fit dimension for both SS and MS events. 
Figure~\ref{fig:source_agree_dnn} shows a comparison of these
two discriminators between simulated and
observed data distributions for the \isot{Ra}{226} calibration source,
and for the measured background-subtracted \twonubb distribution.
While keeping as much discriminating power as possible,
the binning used for each variable is selected to 
minimize systematic errors arising from imperfections
in the MC simulation. 

Since the fit cannot resolve the detailed location of backgrounds 
arising from materials far from the LXe vessel,
the $^{238}$U, $^{232}$Th and $^{60}$Co contributions from such components 
are assigned to fewer representative locations. 
For example, all far $^{238}$U is represented by the decays in 
the air gap between the cryostats and the lead 
shielding in the background model. To account for the errors introduced by 
this approximation, $^{238}$U simulated in the cryostats is used to represent all 
$^{238}$U from remote locations. This is taken to represent an extreme deviaion from
the more realistic case used in the analysis.
The resulting change in the expected number of events near 
$Q_{\beta\!\beta}$ is taken as the systematic error of the background model. 
This is evaluated to be 4.0\% (4.6\%) in \PI (\PII)
by adding contributions from $^{238}$U, $^{232}$Th and $^{60}$Co in quadrature.
In addition, toy studies were used to find the
average bias in the expected number of events near $Q_{\beta\!\beta}$ 
arising from the measured spectral differences between data and MC for energy,
DNN \nonubb discriminator and standoff distance.
The differences between data and MC for their 
distributions obtained with the $\gamma$ calibration
sources are used to correct the predicted
PDFs, while differences in the background-subtracted \twonubb distribution
are used for $\beta$-like components. The relative differences
between results with toy datasets generated from the corrected PDFs, 
but fit without this correction,
are added in quadrature for all contributors 
and sum to 5.8\% (4.4\%) in \PI (\PII). 
Tab.~\ref{tab:shape-syst} summarizes the contributions to background errors,
including other sources unchanged from previous searches.

\begin{table}[h]
\caption{Summary of systematic error contributions.}
\label{tab:shape-syst}
\begin{tabular}{lcc}
\hline
\hline
Source & \PI & \PII \\
\hline
Background errors &  &  \\
\-\hspace{0.5cm} Spectral shape agreement & 5.8\% & 4.4\% \\
\-\hspace{0.5cm} Background model & 4.0\% & 4.6\% \\
\-\hspace{0.5cm} Other~\cite{EXO_PRL17} & 1.5\% & 1.2\% \\ \cline{2-3}
Total error & ~7.1\% & ~6.5\% \\
\hline 
Signal detection efficiency &  &  \\
\-\hspace{0.5cm} Fiducial volume & 2.8\% & 2.6\% \\
\-\hspace{0.5cm} Partial 3D cut & $<0.4\%$ & $<0.4\%$ \\
\-\hspace{0.5cm} Light-to-charge ratio & 0.9\% & 0.9\% \\
\-\hspace{0.5cm} De-noising mis-rec & - & 1.0\%\\
\-\hspace{0.5cm} Other~\cite{Albert2013} & $<1.0$\% & $<1.0$\% \\ \cline{2-3}
Total error & 3.1\% & 3.1\% \\
\hline
\hline 
\end{tabular}
\end{table}

The measured rate of \Rn decays in the LXe is used to constrain 
the appropriate background components arising from these atoms, 
as described in~\cite{Albert2013}. 
The relative rate of cosmogenically produced backgrounds is also constrained~\cite{EXO_cosmogenics}.
In addition, a possible difference between the energy scale from
$\gamma$ calibration sources ($E_\gamma$) and from single- or double-beta decays ($E_\beta$) 
is accounted for by a factor ($B$) that 
scales the energy of the $\beta$-like components in the ML fit, $E_\beta = B E_\gamma$. 
$B$ 
is allowed to freely float and
found to be consistent with unity to the subpercent level in both phases.

The 90\% CL median sensitivity for this \nonubb search
with the DNN \nonubb discriminator is evaluated to be $5.0\cdot 10^{25}$~yr.
The coverage is validated with toy MC studies and 
found to agree with Wilks's theorem~\cite{wilks1938,cowan1998statistical}.
A secondary analysis is performed using a boosted decision trees (BDT) discriminator for MS events
and the BDT discriminator designed in~\cite{EXO_PRL17} for SS events 
as the second fit dimension. 
The BDT for MS is built on variables containing information on 
the energy fraction of the most energetic deposit, 
the spatial spread among deposits and the number of deposits.
The BDT analysis provides comparable 
but slightly worse ($\sim$3\%) sensitivity, suggesting 
that the discrimination power of the DNN discriminator can 
be mostly accounted for by careful construction of BDT variables. 
The DNN analysis was selected as the primary analysis 
prior to unblinding since it had the best sensitivity.

After unblinding the dataset, 
the SS candidate events within $Q_{\beta\!\beta}\pm2\sigma$ were examined, 
which led us to find one 
event, originally with energy in this region, was mis-reconstructed by 
the de-noising algorithm. Its impact on the \nonubb detection efficiency is investigated,
and found to only affect \PII with an efficiency loss $<1.0\%$.
A conservative 1.0\% error is also added to the signal detection efficiency.

\begin{figure*}[bt!]
\centering \includegraphics[width=\linewidth]{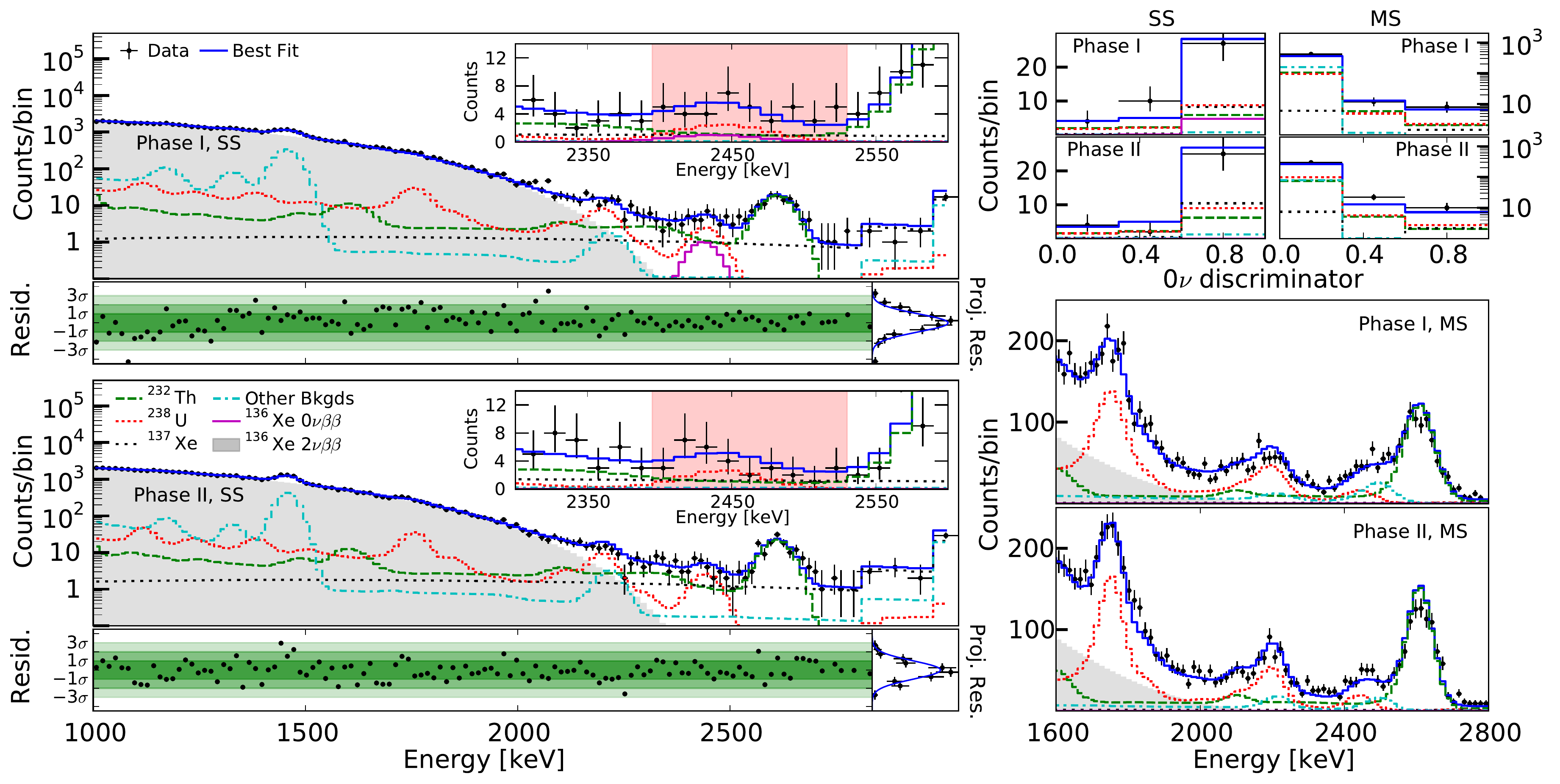}
\caption{
  Best fit to the low background data SS energy spectrum
  for \PI (top left) and \PII (bottom left). The energy bins are 15~keV and 30~keV 
  below and above 2800~keV, respectively.
  The inset shows a zoomed in view 
  around the best-fit value for $Q_{\beta\!\beta}$.  
  (top right) Projection of events  
  in the range 2395~keV to 2530~keV
  on the DNN fit dimension for SS and MS events. (bottom right) MS energy spectra.
  The best-fit residuals typically follow normal distributions, with small deviations taken into 
  account in the spectral shape systematic errors.
}
\label{fig:energy_fit}
\end{figure*}

ML fits are performed to \PI and \PII separately, and the best-fit results 
are shown in Figure~\ref{fig:energy_fit}.
No statistically significant evidence for \nonubb is observed. 
The best-fit background contributions to $Q_{\beta\!\beta}\pm2\sigma$
are shown in Tab.~\ref{tab:roi-bkg}. The rate normalized over the total fiducial Xe mass, 
including all isotopes, is
$(1.7\pm0.2)\cdot10^{-3}$~kg$^{-1}$yr$^{-1}$keV$^{-1}$ and 
$(1.9\pm0.2)\cdot10^{-3}$~kg$^{-1}$yr$^{-1}$keV$^{-1}$
for \PI and \PII respectively. 
The lower limit on the $^{136}$Xe \nonubb half-life is derived 
by profiling over all nuisance parameters,
and results in $T_{1/2}>1.7\cdot10^{25}$~yr ($T_{1/2}>4.3\cdot10^{25}$~yr) 
at 90\% CL in \PI (\PII), while 
the combined limit is $T_{1/2}>3.5\cdot10^{25}$~yr.
This corresponds to an upper limit on the Majorana neutrino mass of
$\langle m_{\beta\!\beta} \rangle < (93-286)$~meV~\cite{DellOro:2016tmg},
using the nuclear matrix elements of~\cite{ibm2:2015,Vaquero:2014dna,QRPA:2014,Menendez2009139,SkyrmeQRPA:2013} 
and phase space factor from~\cite{Kotila:2012zza}.

\begin{table}[h]
\caption{Best-fit background contributions to $Q_{\beta\!\beta}\pm2\sigma$ versus observed number of events in data.}
\label{tab:roi-bkg}
\begin{tabular}{l|cccc|c}
\hline
\hline
(counts) & $^{238}$U & $^{232}$Th & $^{137}$Xe & Total & Data \Tstrut \\
\hline
\PI     & 12.6 & 10.0 & 8.7 & 32.3$\pm$2.3 & 39 \\
\PII    & 12.0 & 8.2  & 9.3 & 30.9$\pm$2.4 & 26 \\
\hline
\hline
\end{tabular}
\end{table}

EXO-200 has concluded its operations reaching a sensitivity 
to Majorana neutrino mass of $78-239$~meV, similar to the most
sensitive searches for \nonubb to date~\cite{KamLAND-Zen:2016pfg,MD_17,Gerda2018,CUORE_First}.
The analysis presented here utilizes a DNN, which maximally makes use of
detailed event topology information for background rejection, leading to a $\sim$25\%
improvement relative to the sensitivity using only event energy and simple SS/MS discriminators.
This performance results from the unique capabilities of a monolithic LXe TPC,
which includes good energy resolution, near maximal signal detection efficiency and 
strong topological discrimination of backgrounds.
This combination holds promise for nEXO~\cite{nEXOpCDR,nEXO-sens},
the planned tonne-scale successor to EXO-200, designed to 
achieve a sensitivity to \nonubb half-life of $\sim$$10^{28}$~yr in \xeiso.

\begin{acknowledgments}
EXO-200 is supported by DOE and NSF in the United States, NSERC in
Canada, SNF in Switzerland, IBS in Korea, RFBR (18-02-00550) in Russia, DFG in
Germany, and CAS and ISTCP in China. EXO-200 data analysis and
simulation uses resources of the National Energy Research Scientific
Computing Center (NERSC).  We gratefully acknowledge the KARMEN
collaboration for supplying the cosmic-ray veto detectors, and the
WIPP for their hospitality.
\end{acknowledgments}

\bibliography{exo_PRL_0nubb}

\end{document}